\documentclass[prb,twocolumn,amsmath,amssymb]{revtex4}
\pdfoutput=1
\usepackage{graphicx}
\usepackage{dcolumn}
\usepackage{bm}

\begin{document}

\newcommand{\up}{\uparrow}
\newcommand{\down}{\downarrow}
\newcommand{\phdagger}{\phantom{\dagger}}          
\newcommand{\aop}[1]{a^{\phdagger}_{#1}}           
\newcommand{\adop}[1]{a^{\dagger}_{#1}}            
\newcommand{\bop}[2]{b^{\phdagger}_{#1}(#2)}           
\newcommand{\bdop}[2]{b^{\dagger}_{#1}(#2)}            
\newcommand{\cop}[1]{c^{\phdagger}_{#1}}           
\newcommand{\cdop}[1]{c^{\dagger}_{#1}}
\newcommand{\dop}[2]{d^{\phdagger}_{#1}(#2)}           
\newcommand{\ddop}[2]{d^{\dagger}_{#1}(#2)}            
\newcommand{\sS}{{\mathcal{S}}}
\newcommand{\sH}{{\mathcal{H}}}
\newcommand{\mgt}{>>}
\newcommand{\mlt}{<<}
\newcommand{\half}{\frac{1}{2}}
\newcommand{\rp}{r^{\prime}}
\newcommand{\om}{\omega}
\def\sO{{\mathcal{O}}}
\def\cd{c^{\dagger}}
\newcommand{\ket}[1]{| #1 \rangle}
\newcommand{\barpsi}[1]{\bar{\Psi}_{#1}}
\newcommand{\psiop}[1]{\Psi_{#1}}

\title{Fermi momentum resolved charge order for spin disordered stripes.}
\author{Mats Granath}
\email{mats.granath@physics.gu.se}
\affiliation{G\"oteborg University\\
G\"oteborg 41296 \\
Sweden
}
\date{\today}

\begin{abstract}
For a doped antiferromagnet with short-range spin stripe correlations and long-range charge stripe order we
find that the manifestation of charge order changes abruptly as a function of momentum along the Fermi surface. 
The disorder averaged local density of states is almost perfectly 
homogeneous when 
integrated only over states which contribute to the ``nodal'' spectral weight whereas it displays long range charge 
stripe order when integrated 
only over states which contribute to the ``antinodal'' spectral weight.  
An effectively two dimensional isotropic nodal liquid can thus coexist with static charge stripes provided there is no static spin order.
We also study commensurate spin and charge stripe ordered systems where the Fermi surface consists of a nodal hole pocket and an open 
 ``stripe band'' section. Due to the stripe order the relation between 
hole density and size of a pocket will be reduced compared to a paramagnet by a factor of two for even charge period and four for odd charge period and
we find an estimated upper limit on the area fraction of a hole pocket of 1.6\% for charge period four and 4\% for charge period five. 
We also discuss why electron pockets are not expected for a stripe ordered system and show that the open Fermi surface section may be electron like with a 
negative Hall coefficient. 
\end{abstract}

\maketitle

\section{Introduction}
Lately it has become increasingly clear that an intrinsic property of the cuprate high-temperature superconductors is the presence of
nanoscale inhomogeneity of the low-energy electronic spectral weight in the doped CuO$_2$ planes. In particular this inhomogeneity seems to come in the form 
of stripes which are unidirectional spin and charge density modulations. From neutron scattering experiments for example there is evidence for a universal spin response
in the hole doped materials that is consistent with dynamic spin stripe correlations.\cite{Tranquada_review} The tendency for stripe formation in a doped 
antiferromagnet is well established and can be understood naively as an effect of the competition between strong antiferromagnetic correlations and 
delocalization of the doped holes.\cite{Machida,Zaanen_Gunnarsson,Emery,Steve_review}

Concurrently with the increasing evidence for stripes there is also a picture emerging in which there are electronic quasiparticles 
in the normal state and Bogoliubov quasiparticles in the superconducting state in line with that expected for a Fermi liquid or a BCS 
superconductor.\cite{ZX_nodal,Davis_interference}
In particular, for underdoped materials these quasiparticles appear to be confined in momentum space to the so called nodal region of the Brillouin zone, near momentum ($\pi/2,\pi/2$), 
where the d-wave gap has a node in the superconducting state. 
The recent observations by scanning tunneling spectroscopy (STS) of a  ``stripe glass'' in which there are static unidirectional 
electronic domains makes the problem even more intriguing.\cite{Davis_stripes}  In fact, this suggests that 
the two phenomena of stripes and nodal quasiparticles can coexist.
A most basic question is how well defined two dimensional quasiparticles can exist in a 
system where there is one dimensional charge order.

In this paper we consider a mean-field model of a system with long-range charge-stripe order and short-range 
spin-stripe correlations that we suggest may hold some clues to this puzzle. From Landau theory describing the low-energy long-distance physics it is 
known that spin stripe order 
always implies charge stripe order but not the converse\cite{Zachar,comment_on_stripes} thus providing the logical background for the study of such a system. 
Assuming translational symmetry along the stripe direction allows us to define a momentum resolved local density 
of states (LDOS) which is integrated only over states with a particular longitudinal momentum. Although this may not be an experimentally measurable quantity it
will give some essential insights into the significance of the lack of long-range spin order.

What we find is an abrupt change in the manifestation of the charge stripe order at low energies as a function of the momentum along the Fermi 
surface. Quite strikingly, as shown in Figure \ref{disorderedLDOS}, the spin disorder averaged LDOS at the Fermi energy when integrated 
only over the states which contribute to the nodal region spectral weight show very little evidence of the charge order and is nearly homogeneous and isotropic, 
while the full LDOS including also antinodal, near ($\pi,0$), states clearly display charge order. This should be contrasted with results for a system with 
both charge and spin order, as shown in Figure \ref{orderedLDOS}, for which the LDOS has a clear signature of charge order for antinodal as well as nodal states.
The results thus suggest that an effectively two dimensional, isotropic nodal state can indeed coexist with static charge stripe order provided that there is no 
static spin order. 
We have considered long range charge order but the results are expected to apply equally well for a system with only local stripe correlations 
provided that the charge order is effectively static and the spin fluctuations are fast compared to the experimental probe. 
 
Another basic question concerns the nature of the Fermi surface of the underdoped cuprates. Angle resolved photoemission (ARPES) experiments observe the famous nodal region 
``Fermi arc'' which is most readily consistent with a 
large Fermi surface closed around the $(\pi,\pi)$ point but where the spectral weight is suppressed by the so called ``pseudogap'' in the antinodal 
region.\cite{ZX_nodal,Valla,ZX_gap,Kanigel,Norman} The 
large Fermi surface picture has been challenged by the recent observations of high frequency quantum oscillations of the Hall resistivity in strong magnetic fields in 
YBa$_2$Cu$_3$O$_y$ (YBCO)
that 
seem to imply small Fermi pockets.\cite{SdH,SdHII,electron_pockets} We show that exactly this distinction is found 
between a system with long-range spin stripe order and a system with only short-range spin stripe correlations. In simulations of commensurate stripe ordered 
systems (Fig. \ref{pockets}) we find  a nodal region
hole pocket together with a quasi-one dimensional Fermi surface section that are replaced by a large Fermi surface in a disordered stripe 
system (Fig. \ref{disorderedLDOS}). 
It is known that a c-axis magnetic field can enhance stripe order\cite{Lake,Tranquada_review}, 
thus providing a possible connection to the observations of quantum oscillations. If the high-field measurements indeed probe a stripe ordered state which the 
ARPES measurements generally do not our results may thus provide an explanation for the apparent discrepancy between the two probes where the Fermi arc could be 
regarded as the remnant in the disordered state of a nodal hole pocket. 

In light of this we also consider the implications of the enlarged stripe unit cell and find that the hole density of a hole pocket will be decreased by a
factor of two for even charge period and four for odd charge period compared to an estimate that assumes no long-range order. Based only on the stripe 
periodicity we can also set an estimated upper limit to the size of a nodal hole pocket of 1.6\% of the full Brillouin zone for a period four stripe and 
4\% for a period five stripe with larger pockets merging 
into open sections. 
From the canonical relation between doping and stripe periodicity\cite{Tranquada_review} we may expect period five for ``1/10'' doping and period four for ``1/8'' doping. 
Quantum oscillations have been observed for doping close 1/10 whereas for 1/8 doping the high field limit remains to be 
explored.\cite{electron_pockets,comment_on_YBCO}
For the period four stripes we expect either no oscillations, corresponding to only open orbits, or possibly a smaller frequency of up to 450T, corresponding to the maximum pocket size.
If a sharp distinction is found between 1/8 and 1/10 it would be a dramatic confirmation of stripe order.

Although hole pockets may explain the frequency of quantum oscillations they cannot explain why the Hall coefficient (R$_H$) may be negative at low 
temperature.\cite{SdH,electron_pockets,Adachi} 
This 
has lead to suggestions for the formation of electron pockets due to broken translational symmetry.\cite{Millis,Chakravarty} However, the location of such electron 
pockets are in regions of the Brillouin zone
(along $(0,0)$ to $(\pi,0)$) where there is no evidence from photoemission of any substantial spectral weight and they must be considered highly speculative. 
For a stripe ordered system, in particular, we argue that electron pockets arise only as an artifact of a mean-field type description 
that effectively ignores the interactions on a stripe. 

Here, instead, we show that the stripe band with open orbits may be electron-like thus providing an alternative 
explanation for a negative Hall coefficient. The sign of R$_H$ for the stripe band depends sensitively on the changes of the Fermi velocity over the Fermi surface which depends in turn 
on the band structure parameters as well as the character (strength and periodicity) of the stripe order.  
The apparent nonuniversality of the Hall coefficient\cite{SdH,electron_pockets,Adachi,Noda,Balakirev} in different materials is not unexpected if stripe order plays a role.

\section{The model\label{model}} 
The model we consider is a tight-binding model on a square lattice in a static potential which couples to the local spin density and 
which may or may not have long-range stripe order. We can think of the potential as a strong inhomogeneous magnetic field that is generated self-consistently 
from an interacting model such as the large-U Hubbard model in the Hartree-Fock approximation.\cite{Fazekas}
The Hamiltonian reads
\begin{eqnarray}
H=-&&t\sum_{\langle rr'\rangle\sigma}(c^{\dagger}_{r,\sigma}c_{r'\sigma}
+{\mbox{H.C.}})\nonumber\\
-&&t'\sum_{\langle rr'\rangle'\sigma}(c^{\dagger}_{r,\sigma}c_{r'\sigma}
+{\mbox{H.C.}})\nonumber\\
+&&\sum_{x,y,\sigma}\sigma(-1)^{y}V(x)c^{\dagger}_{x,y,\sigma}
c_{x,y,\sigma} \; ,\label{H}
\end{eqnarray}
where $c_{r,\sigma}$ is the electron destruction operator at site 
$r=(x,y)$ and with spin $\sigma=\pm$. We will also refer to the stripe perpendicular ($x$) and stripe parallel ($y$) directions as 
$x_\perp$ ($k_\perp$) and $x_\parallel$ ($k_\parallel$) respectively.
The hopping is given in a standard fashion where 
$\langle rr'\rangle$ indicates nearest neighbors, $\langle rr'\rangle'$ next-nearest neighbors. We will use energy units such that 
$t=1$ and we take $t'=-0.3$. For the Hubbard model with on-site interaction $U\sum_r n_{r\uparrow}n_{r\downarrow}$ we would identify the quantity 
$\sigma(-1)^{y}V(x)/U$ with a self-consistent spin-density wave. 
(In a stripe ordered state there would also be a charge density wave generated potential of strength $U\delta n/2$, with $\delta n$ the local density variation,
which will in general be quite negligible 
compared to $V$.) 
In this paper we will not 
attempt to solve any microscopic model self-consistently, instead the Hamiltonian provides a simple means of simulating  
a system with stripe correlations by choosing appropriate potentials $V(x)$.  This type of model has been considered before in several publications\cite{Salkola,MG_stairs},
however not to consider systems with only charge stripe order and neither has the local density of states been explored in any detail.  
We will always consider systems which are
less than half-filled thus providing a most simple caricature of a hole-doped cuprate with stripe order or stripe correlations. We also consider only commensurate 
stripe order in which the stripe period is some integer multiple of the lattice spacing. Although fluctuating incommensurate stripe correlations are found experimentally,
stripe ordered states always appear to be commensurate.\cite{Tranquada_review} 

\section{Ordered Stripes}   

\begin{figure}
\includegraphics[width=8cm]{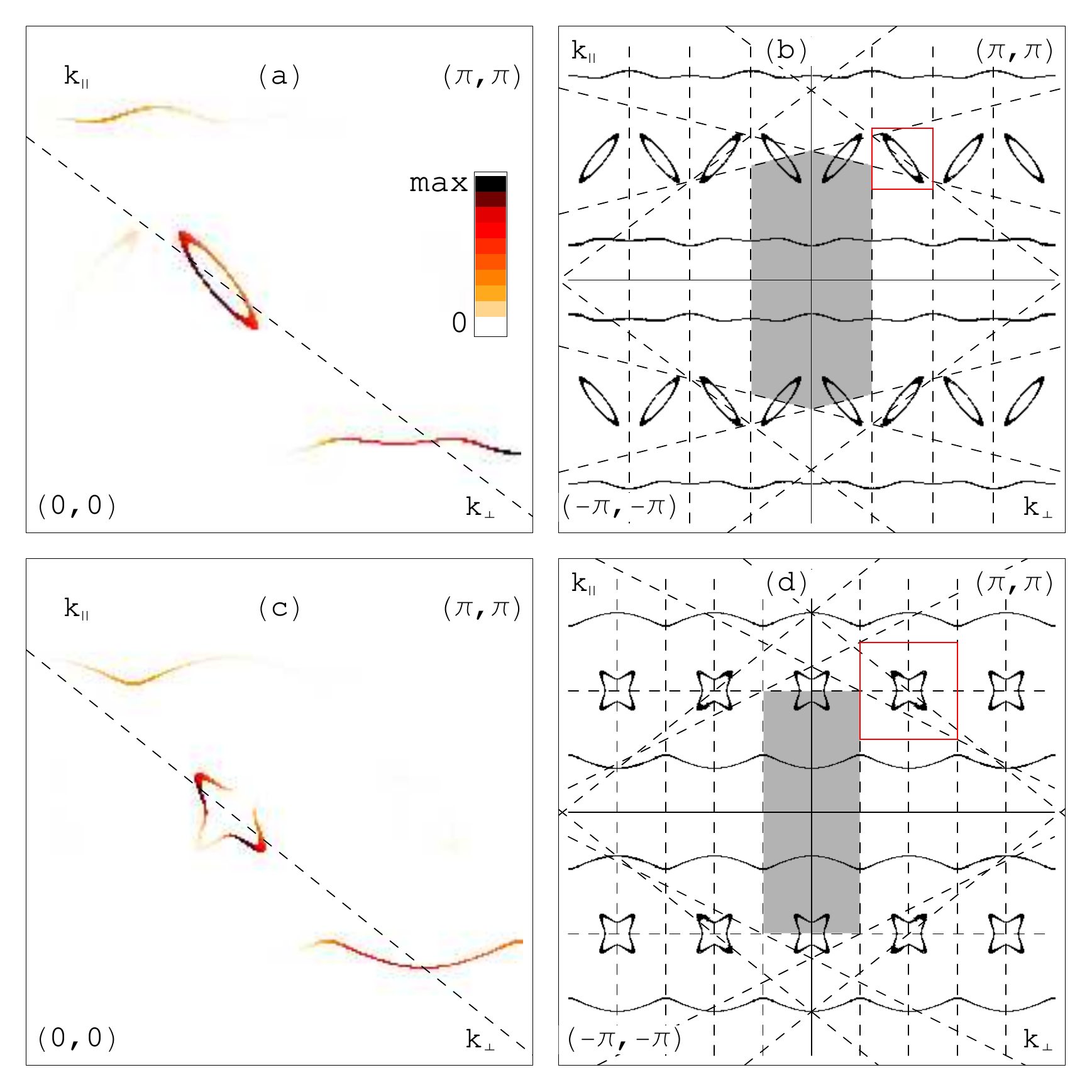}
\caption{\label{pockets}(Color online) Low energy spectral weight of period four (a,b) and period five (c,d) bond-centered stripe ordered systems for stripe potential 
$|V(x)|=0.5$ and doping given below.
The left column shows the spectral weight as an intensity plot where the 
dashed lines are the Bragg planes of the stripe order at $(3\pi/4,\pi)$ and $(4\pi/5,\pi)$ respectively. The right column shows the full Fermi surface together 
with the centermost Bragg planes (dashed) and the first Brillouin zone (highlighted). 
The size of a pocket in b is  
0.7\% of one quadrant ($0<k_\parallel <\pi$, $0<k_\perp<\pi$) and the actual hole density of the nodal pockets is also 0.7\%, with a total hole doping including the antinodal
stripe states of 17.5\% ($E_F=-1.15$). The size of a pocket in 
d is 1.2\% of one quadrant while the hole density of the nodal pockets is only 0.6\%, with a total doping of 12.5\% ($E_F=-1.0$). 
The boxes show the estimated upper limit to the size of a pocket as discussed in the text.}
\end{figure}

Let us first consider the case of a completely regular periodic potential. To exemplify the results we consider bond-centered stripes with charge periodicity four and five. 
For the period four array we choose a potential 
$V_4(x)=0.5V_{4,0}(x)$ 
with $V_{4,0}(x\bmod{8})=1,1,-1,1,-1,-1,1,-1$ and for the period five array  $V_5(x)=0.5V_{5,0}(x)$ with $V_{5,0}(x\bmod{5})=1,1,-1,1,-1$. 
Note the size of the magnetic unit cell 
transverse to the stripes which is twice the charge period only for even periodicities. Considering now also the antiferromagnetic (AF)
$(-1)^y$ modulation (Eq. \ref{H}) along the stripes we have primitive vectors of the magnetic unit cell $(\pi/4,\pm\pi)$ for period four and
$(2\pi/5,0)$ and $(0,\pi)$ for period five. The modulation of the potential is basically the simplest possible consistent with bond-centered stripes.
A self-consistent potential is expected to have a somewhat smaller amplitude on a stripe than between stripes because of the higher hole density but such a 
consideration would not change the qualitative features of our results.

The standard mean field treatment of an antiferromagnet gives a lower and upper Hubbard band separated by an indirect gap 
(for $t'\neq 0$)
while the stripes in the form of $\pi$ phase shifts of the potential transfer some of that weight to midgap states. With moderate doping, 
the chemical potential will lie in this midgap stripe band. Depending on the magnitude of the potential and the doping there may be additional states at the Fermi energy
coming from the stripe modified lower and upper Hubbard bands.     
If the potential is large the lower band will be gapped away from the 
Fermi energy leaving only the midgap stripe states which as we will find have weight primarily in the antinodal region. If the  
potential is small the upper band will also cross the chemical potential, giving rise to electron pockets. For hole doping and considering as a guide 
the large-U Hubbard model such electron pockets are most likely unphysical and we make sure to choose to potential to avoid such states. 
Spectral weight in the nodal region most naturally arises from the lower band but does require a somewhat judicial choice of the magnitude of the potential. 
Whether the stripes are bond-centered or site-centered is however not 
important for the qualitative features of the Fermi surface.   

We diagonalize the Hamiltonian to get a spectrum $\omega_\alpha$ with single-particle eigenstates $\Psi_{\alpha}(\vec{k})$ 
($\alpha$ includes band index and spin) 
that only have spectral weight at momenta connected by multiples of the reciprocal lattice vectors of the stripe unit cell. The 
zero-temperature single-particle spectral function is simply given by $A(\vec{k},\omega)=\sum_{\alpha}|\Psi_{\alpha}(\vec{k})|^2\delta(\omega-\omega_\alpha)$. 
Figure \ref{pockets} shows the Fermi surface (integrated over a narrow window $\delta\omega=0.02$) for the two realizations of stripe potentials. 
For this figure we have chosen the Fermi energy and corresponding hole doping to give a 
simple Fermi surface 
with only two bands contributing, one band giving the nodal Fermi pocket and one band giving the quasi-one dimensional antinodal weight. 
The antinodal spectral weight is qualitatively not very sensitive to the Fermi energy, the Fermi surface only moves along the stripe direction $k_\parallel$ as 
expected for a quasi-one dimensional band. 
The nodal weight on the other hand is very sensitive to the details of the potential and the doping, in particular for higher doping
several bands may contribute and give superimposed nodal hole pockets with different sizes.

\subsection*{Fermi surface}
We draw the following general conclusions from the calculations on ordered stripe arrays: 
(1) The size of a nodal hole pocket does not give the corresponding hole density without an a priori knowledge 
of the stripe order. Naively, for a paramagnet with one pocket per quadrant of the full 1st Brillouin zone, one might infer a hole density which is 
$n_h=8A$ where $A$ is the area fraction of a pocket. Here we find that the actual hole density is one half of this value $n_h=4A$ for period four stripes and only one 
quarter $n_h=2A$ for period five stripes, i.e. there are only two or one pocket respectively in the reduced first Brillouin zone. We have checked that these 
results generalize to longer 
stripe periodicities with the same distinction between even and odd periodicity. Thus, only for even periodicity is the relation between hole density and the 
size of a pocket the same as for an antiferromagnet.\cite{Rice}  
 
(2) The bulk of the hole density for this system is actually in the quasi-one dimensional stripe band which may dominate the charge transport.\cite{Noda,Ando}
However, since the corresponding Fermi surface section does not have any closed orbit this band would not give rise to quantum oscillations. 

(3) The size of the Brillouin zone limits the size of a nodal pocket, larger pockets will merge to form extended Fermi surface sections (open orbits). 
The size limit depends on the shape and orientation of the pocket, however we can make a general and conservative upper estimate by assuming that the extension of the pocket 
is of the same magnitude along both directions as indicated by the boxes in Figure \ref{pockets} (b) and (d). With this we estimate the maximum area fraction 
(of the full extended BZ) to $A_{4,max}=1/64\approx 1.6\%$ for charge period four and $A_{5,max}=1/25=4\%$ for period five. 

The above estimate does appear to rule out a period four stripe ordered state in the system (YBa$_2$Cu$_3$O$_{6.5}$) with hole doping $p=0.10$ in which quantum oscillations were 
recently observed 
with a pocket of size 1.9\%.\cite{SdH} The size is however well within the limit for a period five state, which may be reasonable given the observations by inelastic neutron 
scattering of magnetic correlations that are consistent with this stripe periodicity\cite{Dai,Tranquada_review} 
For charge-period five the size of the pocket would correspond to hole density of only 3.8\%, while the remaining 
6.2\% hole density would according to this model reside 
in the quasi-one dimensional stripe band. 

The cyclotron mass of a pocket is given 
by\cite{Ashcroft}
$\frac{m*}{m_e}=\frac{\hbar^2}{2\pi m_e}\frac{\partial A}{\partial\epsilon}\approx 2.61\frac{\partial \tilde{A}}{\partial\tilde{\epsilon}}$ where we have used 
a lattice spacing of $3.8$\AA, $t=0.4 eV$, $\tilde{A}$ is the area fraction of one quadrant of the Brillouin zone and $\tilde{\epsilon}$ is energy in units of $t$. 
A problem with these pockets in relation
to the experiments is that the dispersion is quite steep which gives a cyclotron mass of around $0.3$ electron masses for both the period four and period 
five stripes which is much smaller than the $1.9$ reported experimentally. However, one may expect that strong correlations can significantly narrow the bands and
increase the cyclotron mass.

\subsection*{Electron pockets and open orbits.}
The Hall resistivity is negative under the conditions in which the quantum oscillations have been detected which from semiclassical transport theory implies 
electron like charge carriers instead of the expected hole like. 
This led to a recent suggestion that 
electron pockets, not hole pockets, in a stripe ordered system may be responsible for the oscillations.\cite{Millis} In contrast to the hole pockets discussed in this paper 
such electron pockets are centered on the stripes and may appear with a suitable choice of the potential.
However, we would argue that they are unphysical as they correspond effectively to occupying states in the upper 
Hubbard band of the stripe, with the relatively low energy being an
artifact of the mean 
field like treatment of the stripe. 

We can see this by considering a single site-centered stripe in the large $U$ limit. Self-consistently this corresponds to a 
large staggered potential outside of the stripe center row giving rise to a decoupled tight binding chain on the stripe.\cite{MG_stairs} The problem with this is 
particularly obvious at half-filling
of the stripe where the procedure gives a gapless electronic band instead of the appropriate insulator with a gap proportional to $U$. This discrepancy is related
to the fact that the mean field treatment cannot accurately describe the effectively paramagnetic state on the site-centered stripe. 
Clearly the naive stripe potential misses essential aspects of the interactions on a stripe, a deficiency that will remain also away from half-filling and for a 
smaller stripe potential.

\begin{figure}
\includegraphics[width=8cm]{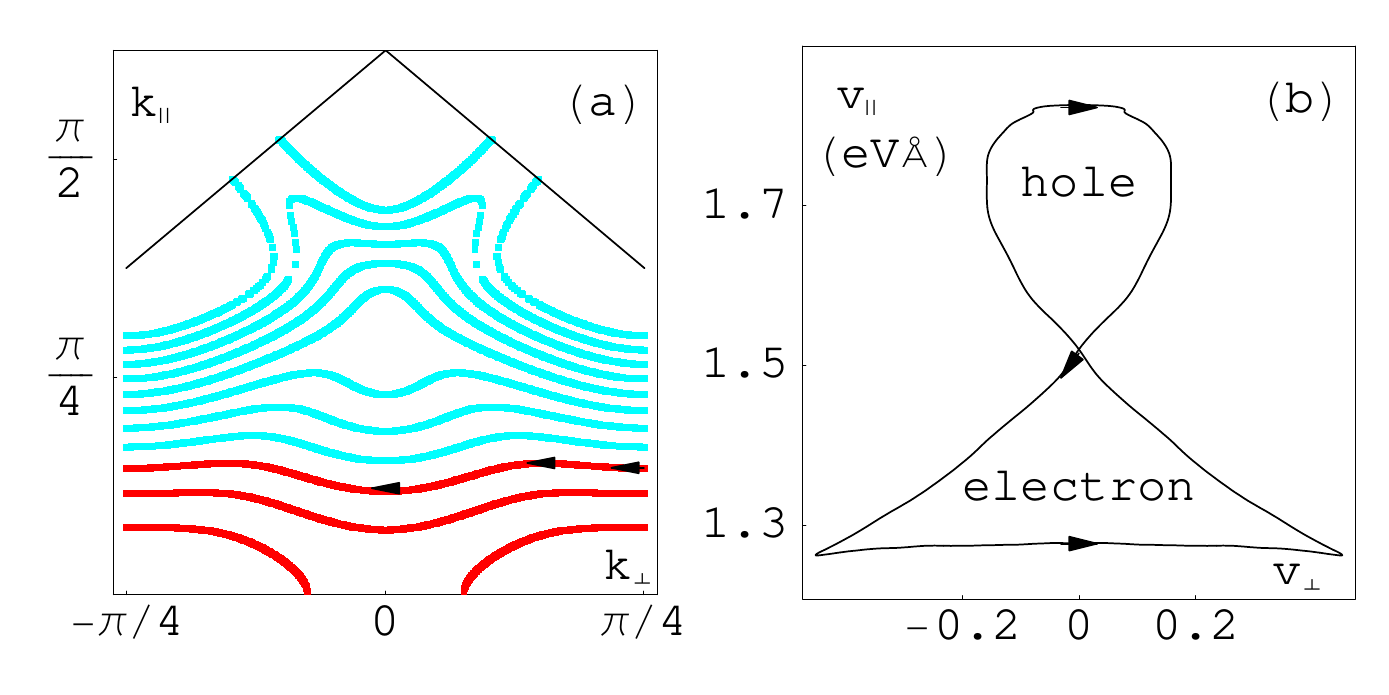}
\caption{\label{orbits}(Color online) Fermi surface sections (a) in the upper 
half of the first BZ of the stripe band for the charge period four stripe (same as in Fig. \ref{pockets} a,b) for varying chemical potential. 
The three lower sections are electron-like and the rest are hole-like. As explained in the text the Hall coefficient is proportional to the area of the integral 
over the Fermi velocity 
curve as shown in (b) for the section in (a) marked by arrows. There is a contribution from both a hole-like and an electron-like loop. The magnitude of the 
longitudinal ($v_\parallel$) and transverse ($v_\perp$) velocities are calculated using a lattice spacing of $3.8${\AA} and hopping $t=0.4 eV$.}
\end{figure}

The resolution to the negative Hall resistivity may lie instead in the charge transport related 
to the open stripe bands\cite{Adachi,Noda} with the oscillations being due to hole pockets. For the transport properties of the quasi-one dimensional stripes it 
may be important to consider correlations\cite{Lubensky} as evidenced by the antinodal pseudogap but here we will do the standard semiclassical calculation. Following the presentation of
Ong\cite{Ong}, the sign of the Hall conductivity is given by the integral over the Fermi surface $\hat{z}\cdot\int d\vec{l}\times\vec{l}$ where 
$\vec{l}=\vec{v}_{\vec{k}}\tau_{\vec{k}}$ is
the scattering length vector with Fermi velocity $\vec{v}_{\vec{k}}$ and relaxation time $\tau_{\vec{k}}$. Figure \ref{orbits}a shows constant energy curves of 
the ``stripe band'' of the 
charge period four system considered earlier (same parameters). With increasing energy these evolve from electron pockets into open sections and into 
hole pockets with the Hall 
coefficient changing continuously from negative (electron-like) to positive (hole-like). (We have assumed a $\vec{k}$ independent relaxation time.) 

Figure  \ref{orbits}b shows the area traced out by Fermi velocities for the particular section marked by arrows as the Fermi surface is followed across the 
Brillouin zone where the velocities $v_\parallel$ and $v_\perp$ corresponding to the longitudinal and transverse components respectively. We see that the outer 
part of the Fermi surface ($k_\perp\approx\pi/4$) gives a hole-like contribution whereas the inner part ($k_\perp\approx 0$) is electron-like.  
The particular Fermi surface displayed in Figure \ref{pockets} has a hole-like open section but is close to the transition between hole-like and electron-like.

The sign of the 
Hall coefficient depends sensitively on the curvature of Fermi surface and the Fermi velocity and small changes can switch the sign. The fact that experimentally the sign
of the Hall coefficient in strong magnetic fields appears to be non-universal may be an indication that it is related to such open orbits.   
A more realistic calculation of the stripe band-structure is needed including also an examination of the total Hall coefficient due to a possibly 
electron-like open 
Fermi surface section together with hole pockets. 

\subsection*{Charge order} 
Let us also consider the local density of states of this system $LDOS(\vec{x},\omega)=\sum_\alpha |\Psi_{\alpha}(\vec{x})|^2\delta(\omega-\omega_\alpha)$. We 
will be interested in considering the connection between the spectral function and the local density of states at the Fermi energy and in particular the distinction between
nodal and antinodal states. Since momentum is conserved modulo $\pi$ along the stripe direction $k_\parallel$ we can associate to every state a momentum 
$-\pi/2<k_{\parallel,\alpha}\leq \pi/2$ and choose to make a partial sum in the LDOS that only 
includes states with certain $k_\parallel$. We define 
\begin{eqnarray}
LDOS_{nodal}&=&\sum_{\alpha:|k_{\parallel,\alpha}|>\pi/4} |\Psi_{\alpha}(\vec{x})|^2\delta(\omega-\omega_\alpha) \nonumber \\ 
LDOS_{a-nodal}&=&\sum_{\alpha:|k_{\parallel,\alpha}|\leq \pi/4} |\Psi_{\alpha}(\vec{x})|^2\delta(\omega-\omega_\alpha) \label{LDOS}
\end{eqnarray} 

\begin{figure}
\includegraphics[width=8cm]{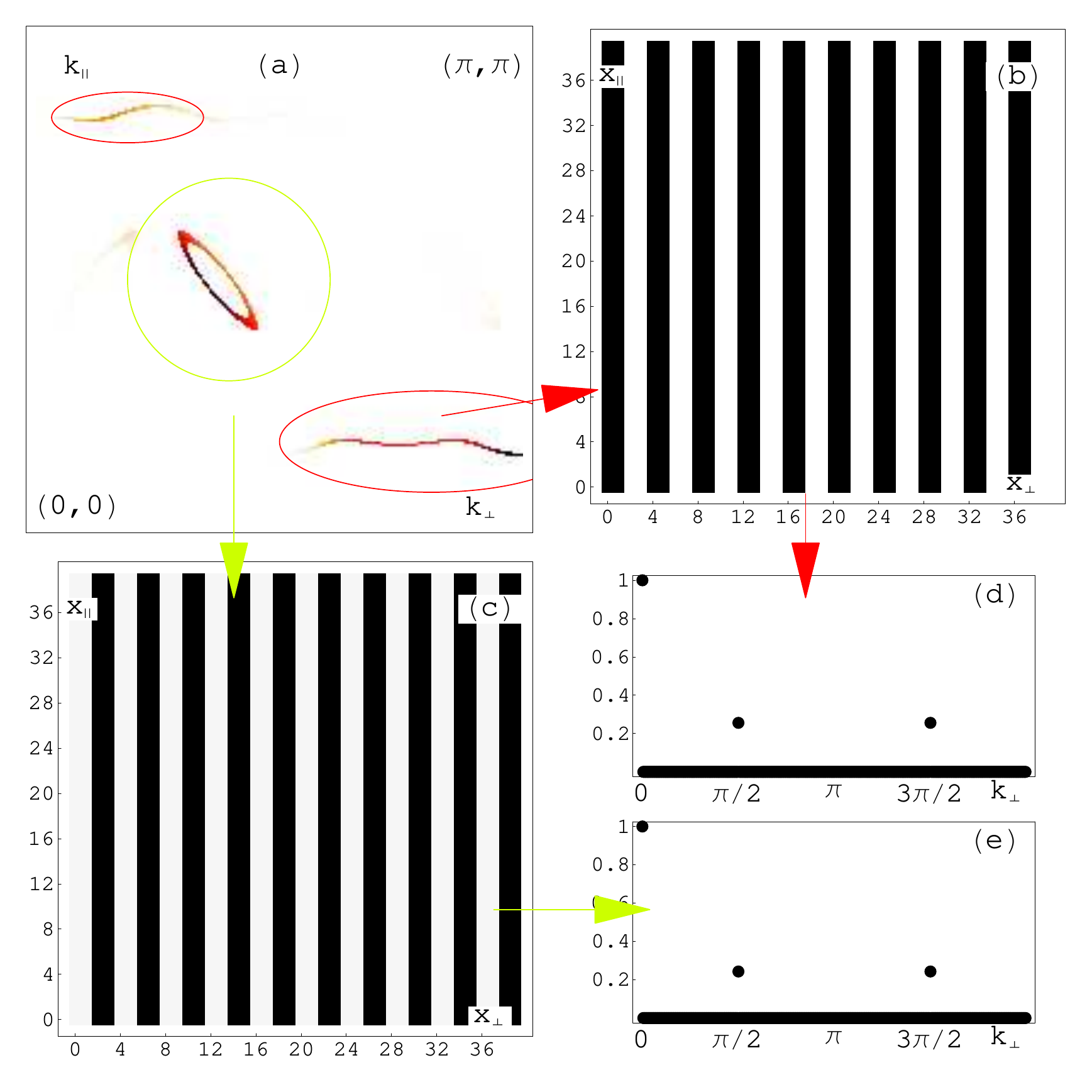}
\caption{\label{orderedLDOS}(Color online) Spectral intensity at the Fermi energy (a) together with the local density of states of the antinodal (b) 
and nodal (c) states plotted on the 
same normalized scale with respect to the  
maximum density (black). 
Note that the high density regions are shifted by two sites between nodal and antinodal weight. The antinodal weight has high density at the AF domain
walls of the stripe potential whereas the nodal weight has high intensity between. Also shown is the amplitude of the Fourier transform of the 
LDOS transverse to the stripe direction with the uniform density component $k_\perp=0$ normalized to one. The LDOS modulations in both regions is roughly 30\% of
the mean density. (The combined total LDOS modulation at the Fermi energy, not displayed, is only around 5\%. The integrated charge modulation over all occupied states is 
around 4\%.) }
\end{figure}

Figure \ref{orderedLDOS} shows the nodal and antinodal LDOS for the period four array, the antinodal states have a higher density on sites $x\bmod{4}=$0 or 1  
while the nodal states have a higher density on sites $x\bmod{4}=$2 or 3 (with respect to the potential $V_{4,0}(x\bmod{8})=1,1,-1,1,-1,-1,1,-1$).
As emphasized in several other publications\cite{Eder,MG_dichotomy} the nodal states derive from the 
antiferromagnet whereas
the antinodal states are the ``stripe states'' living on domain walls of the antiferromagnet. 
We will find that this distinction will be crucial when we in the next section consider a system without long range
spin stripe order.  

\section{Disordered Stripes}

We will now consider a charge ordered but spin disordered system. We generate a stochastic spin-stripe potential with a short correlation length through an update 
procedure $V(x+1)=V_0(x)V_0(x+1)V(x)+\psi$ where $-2m<\psi<2m$ is a random number and apply a cutoff $-m<V(x)<m$ at every step.  
Here $V_0(x)$ gives the sign of the potential for the 
corresponding spin-stripe ordered array which in the case of the simple charge-period four bond-centered stripe 
takes the value $V_0(x\bmod{8})=1,1,-1,1,-1,-1,1,-1$.    
The distribution is readily evaluated analytically and we find a disorder averaged correlation function 
\begin{equation}
\langle V(x)V(x+j)\rangle=\frac{2}{3}m^2V_0(x)V_0(x+j)e^{-\kappa |j|}\,,
\label{correlator}
\end{equation}
with a short correlation length $\xi=1/\kappa=1/\ln(2)\approx 1.44$. We will take $m=1$ which we find gives a good correspondence with the ordered system 
considered 
previously, i.e. the same Fermi energy gives roughly the same doping. 
In order to be able to consider large system sizes (in the transverse direction) we still assume antiferromagnetic order along the stripes. Assuming short-range  
correlations would broaden the states in $k_\parallel$ but as far as the transverse charge order is concerned we do not expect this 
to have any significant impact. For the numerical calculations we use a transverse system size $N$ which is large compared to the correlation length of the potential 
(typically $N=120$) and diagonalize the system for a given static but random configuration of the potential and take the average over many realizations.

Physically what we have in mind is a system with fluctuating short-range spin stripe correlations with some correlation length $\xi$ and 
frequency $\omega_0$.
The disorder average can thus be relevant either for a fast, compared to $\omega_0$, but extended probe such as inelastic neutron scattering or ARPES (disorder average 
in real space) or a slow but local probe such as STS (disorder average in time). There is of course a significant approximation involved in representing the 
spin fluctuations by a sequence of static spin configurations. In particular quasi-particle dynamics will not be accurately represented using such a 
quasi-static spin approximation.
However for the gross and qualitative features of the spectral weight distribution we expect that the approximation can give reasonable results.

\begin{figure}
\includegraphics[width=8cm]{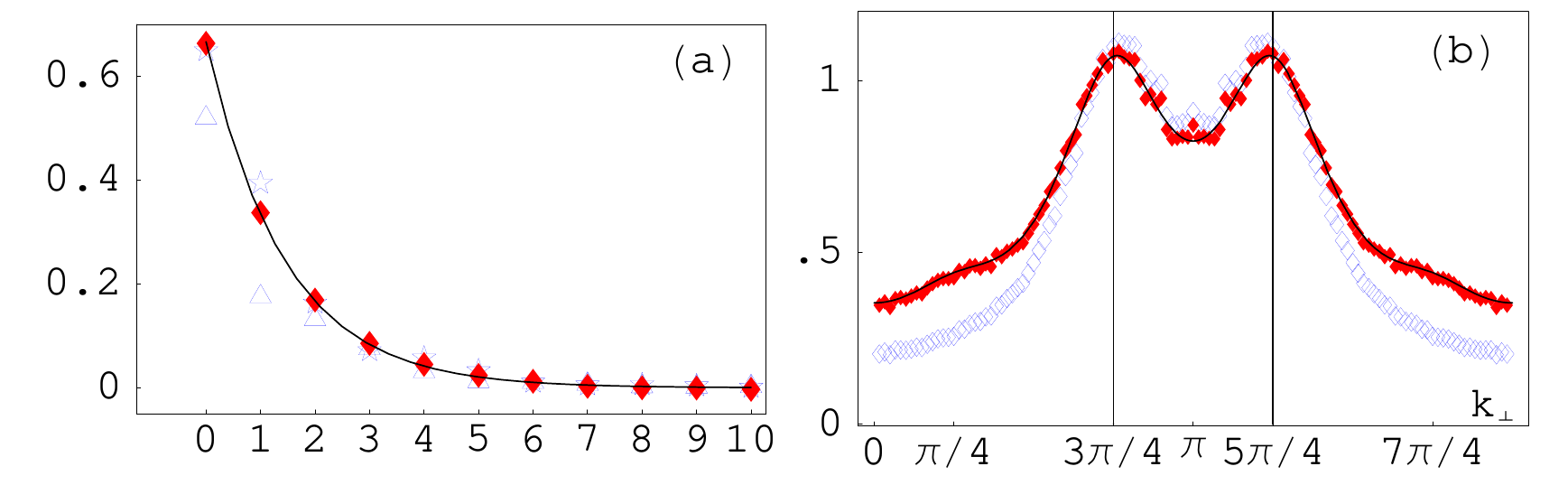}
\caption{\label{potential} (Color online) Disorder averaged spin density correlator transverse to the stripe extension for period four bond-centered stripes showing the 
approximate self-consistency of the stripe potential. 
In (a) the absolute value in real space and in (b) the Fourier 
transform. Solid points are values for the stripe potential and open symbols for the actual spin densities
which follow from the potential. The latter are multiplied by $U^2=(3.5)^2$ as discussed in the text. 
In (a) the $0$ position is on a stripe (triangles) or between stripes (stars), where the smaller amplitude on a stripe is consistent with a higher hole density. 
The lines are analytic results for the disorder averaged potential, Equations \ref{correlator} (absolute value and $m=1$) and \ref{Spot} in 
(a) and (b) respectively. 
The average is taken over 5000 runs with system size 120$\times$120 (unit cell 120$\times$2).}
\end{figure}

\begin{figure}
\includegraphics[width=8cm]{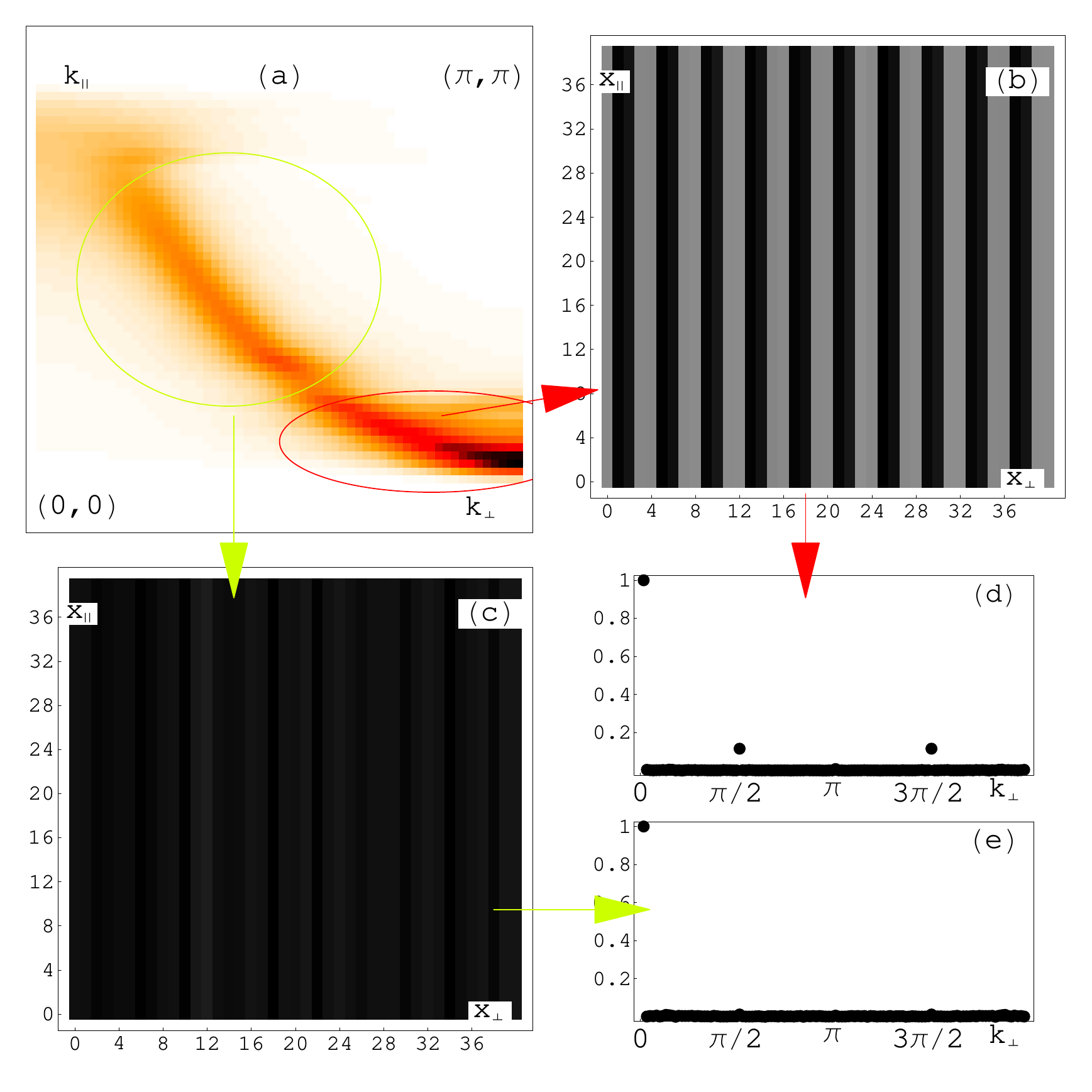}
\caption{\label{disorderedLDOS} (Color online) Spectral intensity (a) of spin disordered system (same as in Fig. \ref{potential}) together with the LDOS at the Fermi energy of antinodal (b) and nodal states (c). The LDOS is shown for an arbitrary section of the system
and is plotted on the same relative scale as for the spin-stripe ordered system (Fig. \ref{orderedLDOS}). The relative variations is around 18\% for the antinodal LDOS and
around 3\% for the nodal LDOS. The components at momenta $\pi/2$ and $3\pi/2$ correspond to roughly 15\% density variations of the antinodal LDOS (d) and 1.5\% 
(barely visible in e) for the nodal LDOS variations with the rest of the variations being due to random disorder. 
The total density modulation integrated over all filled 
states is around 3\%.
The simulation is averaged over 10000 runs with system 
size 120$\times$120.}       
\end{figure}

\subsection*{Spin correlations}  

We think of the stripe potential as reflecting the instantaneous spin configuration. Clearly this disordered potential will not be a self-consistent mean 
field solution to the Hubbard model. Nevertheless, to be physically reasonable we may at least require that the disorder averaged potential should be roughly self 
consistent in the sense that it faithfully reflects the average spin configuration.    
For the potential as we have chosen it every site is equivalent up to signs given by $V_0$, however for the actual disorder averaged spin density we need to 
distinguish 
between two inequivalent sites which are on average on a stripe ($x\bmod{4}=$0 or 1) or between a stripe ($x\bmod{4}=$2 or 3) with respect to the potential 
$V_0(x\bmod{8})=1,1,-1,1,-1,-1,1,-1$. 
Figure \ref{potential} shows the corresponding spin-spin 
correlators $|\langle S^z(x)S^z(x+j)\rangle|$, 
as well as the Fourier transforms $S_{pot}(q)=\langle (1/N)\sum_{x_1,x_2}V(x_1)V(x_2)e^{iq(x_2-x_1)}\rangle$ and 
$S(q)=\langle (1/N)\sum_{x_1,x_2}S^z(x_1)S^z(x_2)e^{iq(x_2-x_1)}\rangle$. From the analytic expression for the correlation function of the potential, 
Eqn. \ref{correlator} with $m=1$, we can calculate 
\begin{eqnarray}
\label{Spot}
S_{pot}(q)&=&\frac{2}{3}\Biglb(-1+\frac{2-\sqrt{2}}{8}\sum_{q_0=\pm \pi/4}\frac{e^{\kappa}-\cos(q-q_0)}{\cosh(\kappa)-\cos(q-q_0)}\nonumber\\
&+&\frac{2+\sqrt{2}}{8}\sum_{q_0=\pm 3\pi/4}\frac{e^{\kappa}-\cos(q-q_0)}{\cosh(\kappa)-\cos(q-q_0)}\Bigrb)  
\end{eqnarray}
with $\kappa=\ln(2)$.
The spin-spin correlators are multiplied by a factor $U^2=(3.5)^2$ and the good agreement with the potential indicates that the averaged potential
is quite close to being self-consistent for the Hubbard model with $U=3.5$. The main source of the discrepancy 
is caused by the fact that the potential assumes on average a uniform magnitude of the local magnetization whereas a self-consistent calculation gives a smaller 
magnetization on the stripes (Fig. \ref{potential}a) consistent with a larger hole density. This can be fine tuned without any significant impact on the subsequent results
and we will prefer to keep the model as simple as possible.

\subsection*{Charge order}
Let us now study the charge density in this system with short range spin correlations. Here we consider two objects, the disorder averaged local density of states 
$\tilde{LDOS}(\vec{x},\omega)=\langle LDOS(\vec{x},\omega)\rangle$ and the disorder averaged spectral function 
$\tilde{A}(\vec{k},\omega)=\langle A(\vec{k},\omega)\rangle$ where we can again divide the LDOS into nodal and antinodal parts according to Eqn. \ref{LDOS}. 
 
Figure \ref{disorderedLDOS} shows the spectral weight integrated over a window ($\delta w=0.05$)\cite{window} close to the Fermi energy, as well as the corresponding 
nodal and anti-nodal LDOS. We take the Fermi energy the same ($E_F=-1.15$) as for the period four stripe ordered system considered above which corresponds here to an 
average hole doping of around 18\%. 
Due to the disorder the spectral weight is spread out such that the stripe Fermi surface and nodal pocket loose their integrity and
merge into a single large Fermi surface which if we symmetrize with respect to stripes along both directions would close around the $(\pi,\pi)$ point.

What is particularly notable are the results for the LDOS where the nodal LDOS is almost uniform with only
a very weak signature of the charge order
 whereas the antinodal LDOS 
shows long range charge-stripe order not very different in magnitude from that of the stripe ordered system considered previously. For the nodal states there is an apparent 
restoration of the rotational symmetry of the lattice for the spin disordered system. The results can be roughly understood based on the knowledge of the 
stripe ordered system in which the antinodal weight comes from stripe states that are centered on antiferromagnetic 
domain walls and are quite insensitive to whether or not there is long-range and static spin order. The nodal states on the other hand live 
primarily between stripes for a spin-stripe ordered state and the spin fluctuations apparently conceal the stripes. 
A complementary understanding can be found from earlier work on the same model where it was noted that in the presence of weak potential disorder the 
localization length transverse to the stripes of the antinodal states is of 
the order of or less than the stripe period whereas the localization length of the nodal states can be several times the stripe period.\cite{MG_dichotomy}

\section{Discussion}

The Fermi momentum resolved charge order found here is  
yet another manifestation of what has been called the ``nodal-antinodal dichotomy'' in which the low-energy spectral properties 
are very different for the nodal and antinodal regions.\cite{ZX_nodal}
In particular the enigmatic pseudogap is known from ARPES to be exactly such a momentum resolved phenomenon 
in which there is a partial suppression of the antinodal spectral weight.

Let us explore the implications of an antinodal pseudogap on our results in a qualitative fashion: First, it would lift the bulk of the antinodal 
spectral weight away from the Fermi energy leaving only the nodal Fermi arc.
In addition it would imply that the charge-stripe order may only be visible above the pseudogap energy scale. 
The latter property does indeed seem to be realized for the ``stripe glass'' observed by tunneling spectroscopy 
for which the differential tunneling conductance is nearly uniform at low energies and only show the unidirectional modulations at energies around the 
pseudogap energy.\cite{Davis,Davis_stripes}  
These observations together with the results presented here give a very strong indication that the pseudogap is in fact related to stripes.
Also corroborating this picture is ARPES data which 
find that the pseudogap disappears abruptly with the momentum along the Fermi surface in going from $(\pi,0)$ to $(\pi/2,\pi/2)$\cite{ZX_gap,Kanigel,KanigelII} 
in a fashion analogous to the charge stripe
order presented here.

The pseudogap is then a strong correlation gap (spin-gap) on a single stripe characterized as an interacting quasi-one 
dimensional electron gas.\cite{Zachar_spingap,Steve_review} 
In the work presented here any strong correlation effects
apart from the stripe order itself are completely ignored but we do find that the antinodal spectral weight comes from the quasi-one dimensional stripe 
states thus providing a logical consistency with an antinodal pseudogap from stripes.
One way to argue why it may be reasonable to treat the stripe correlations separately is if there is a separation of 
energy scales in the renormalization group sense in which the antiferromagnetic correlations, 
which for a doped system correspond to stripe correlations, dominate and 
residual correlations which gives rise to the pseudogap become prominent at a lower energy scale.\cite{Erica,MG_nodal,Steve_review} 

As is well known for the underdoped cuprates 
the superconducting order occurs at an even lower energy scale.
From, for example, ARPES results on the gap evolution\cite{ZX_gap,KanigelII} and STS results on the Bogoliubov quasiparticle interference\cite{Davis_interference} 
it is by now quite clear that long-range superconducting order is intimately linked to the nodal liquid which as we have found can be isotropic even in the presence of
charge stripe order. In this context we note that the spin stripe 
ordered La$_{1.875}$Ba$_{0.125}$CuO$_4$ has a significantly suppressed superconducting transition temperature while there is still a large antinodal 
pseudogap.\cite{Valla,comment_on_Valla} 
The suppression of superconductivity in this system may thus be related to anisotropy of the nodal liquid due to spin stripe order.\cite{Li}

As a continuation of this work the modeling of disordered stripes should be extended to include also some characterization of the pseudogap and
a d-wave like pair-field which could allow for a more detailed comparison with STS and ARPES results. 
For the ordered stripe phases work is in progress to calculate the total conductivity resulting from the contributions of hole 
pocket and open stripe band and to characterize more generally the sign and oscillations of the Hall resistivity in such a system. 

 
To summarize, we find that for a charge ordered but spin disordered stripe state there is a natural separation into nodal and antinodal states for which the latter shows
a much stronger signature of the charge order. The local density of states at the Fermi energy is almost homogeneous when integrated only over states that contribute
to the nodal region spectral weight. We also consider commensurate spin and charge stripe ordered systems where the Fermi surface consists of a nodal hole pocket and an 
open (quasi-one dimensional) antinodal section. The relation between the size of the pocket and the hole density depends on whether the charge periodicity is odd or even 
and we give upper estimates of the possible size of a pocket for charge period four and five of 1.6\% and 4\% respectively. We find that the open section may be electron or hole
like depending on the details of the stripe band structure.

We thank John Tranquada and Steven Kivelson for comments and S\'eamus Davis for discussions of his experimental data\cite{Davis}.


\begin{thebibliography}{12}
\bibitem{comment_on_stripes} Here we will only consider stripes which are antiphase domain walls of the antiferromagnetic order such that the ordering vector of the charge
is twice that of the spin, $\vec{k}_{ch}=2\vec{k}_{spin}$
\bibitem{comment_on_YBCO} Here we refer to the estimated hole concentrations for the ortho-II ($p=0.10$) and ortho-VIII ($p=0.12$) phases of 
YBa$_2$Cu$_3$O$_y$ in Ref. \onlinecite{electron_pockets}.
\bibitem{window} Here we take a slightly larger integration window than for the ordered arrays to get a good disorder average.
\bibitem{comment_on_Valla} The lack of any direct evidence of the stripe 
order (such as Fermi surface pockets) in the ARPES data\cite{Valla} is presumably due to the finite correlation length of the stripe order.
\end{thebibliography}
\end{document}